\documentclass[twocolumn]{aastex631}

\usepackage{xspace}

\newcommand{\um}{\ensuremath{\mu\rm{m}}\xspace}
\newcommand{\kms}{\ensuremath{\rm{km\,s}^{-1}}\xspace}

\newcommand{\uJy}{\ensuremath{\mu\rm{Jy}}\xspace}

\newcommand{\cii}{[C{\scriptsize II}]\xspace}

\usepackage[encapsulated]{CJK}

\begin{document}

\title{A confirmed deficit of hot and cold dust emission in the most luminous Little Red Dots}

\shorttitle{Luminous LRDs are FIR faint}
\shortauthors{Setton et al.}

\author[0000-0003-4075-7393]{David J. Setton}\thanks{Email: davidsetton@princeton.edu}\thanks{Brinson Prize Fellow}
\affiliation{Department of Astrophysical Sciences, Princeton University, 4 Ivy Lane, Princeton, NJ 08544, USA}

\author[0000-0002-5612-3427]{Jenny E. Greene}
\affiliation{Department of Astrophysical Sciences, Princeton University, 4 Ivy Lane, Princeton, NJ 08544, USA}

\author[0000-0003-3256-5615]{Justin~S.~Spilker}
\affiliation{Department of Physics and Astronomy and George P. and Cynthia Woods Mitchell Institute for Fundamental Physics and Astronomy, Texas A\&M University, 4242 TAMU, College Station, TX 77843-4242, US}

\author[0000-0003-2919-7495]{Christina C. Williams}
\affiliation{NSF's National Optical-Infrared Astronomy Research Laboratory, Tucson, AZ 85719, USA}
\affiliation{Steward Observatory, University of Arizona, 933 North Cherry Avenue, Tucson, AZ 85721, USA}

\author[0000-0002-2057-5376]{Ivo Labb\'e}
\affiliation{Centre for Astrophysics and Supercomputing, Swinburne University of Technology, Melbourne, VIC 3122, Australia}

\author[0000-0002-0463-9528]{Yilun Ma  (\begin{CJK*}{UTF8}{gbsn}马逸伦\ignorespacesafterend\end{CJK*})}
\affiliation{Department of Astrophysical Sciences, Princeton University, 4 Ivy Lane, Princeton, NJ 08544, USA}

\author[0000-0001-9269-5046]{Bingjie Wang (\begin{CJK*}{UTF8}{gbsn}王冰洁\ignorespacesafterend\end{CJK*})}
\affiliation{Department of Astronomy \& Astrophysics, The Pennsylvania State University, University Park, PA 16802, USA}
\affiliation{Institute for Computational \& Data Sciences, The Pennsylvania State University, University Park, PA 16802, USA}
\affiliation{Institute for Gravitation and the Cosmos, The Pennsylvania State University, University Park, PA 16802, USA}

\author[0000-0001-7160-3632]{Katherine E. Whitaker}
\affiliation{Department of Astronomy, University of Massachusetts, Amherst, MA 01003, USA}
\affiliation{Cosmic Dawn Center (DAWN), Denmark}

\author[0000-0001-6755-1315]{Joel Leja}
\affiliation{Department of Astronomy \& Astrophysics, The Pennsylvania State University, University Park, PA 16802, USA}
\affiliation{Institute for Computational \& Data Sciences, The Pennsylvania State University, University Park, PA 16802, USA}
\affiliation{Institute for Gravitation and the Cosmos, The Pennsylvania State University, University Park, PA 16802, USA}

\author[0000-0002-2380-9801]{Anna de Graaff}
\affiliation{Max-Planck-Institut f\"ur Astronomie, K\"onigstuhl 17, D-69117, Heidelberg, Germany}

\author[0000-0002-8909-8782]{Stacey Alberts}
\affiliation{Steward Observatory, University of Arizona, 933 North Cherry Avenue, Tucson, AZ 85721, USA}

\author[0000-0001-5063-8254]{Rachel Bezanson}
\affiliation{Department of Physics and Astronomy and PITT PACC, University of Pittsburgh, Pittsburgh, PA 15260, USA}

\author[0000-0002-3952-8588]{Leindert A. Boogaard}
\affiliation{Leiden Observatory, Leiden University, PO Box 9513, NL-2300 RA Leiden, The Netherlands}

\author[0000-0003-2680-005X]{Gabriel Brammer}
\affiliation{Cosmic Dawn Center (DAWN), Niels Bohr Institute, University of Copenhagen, Jagtvej 128, K{\o}benhavn N, DK-2200, Denmark}

\author[0000-0002-7031-2865]{Sam E. Cutler}
\affiliation{Department of Astronomy, University of Massachusetts, Amherst, MA 01003, USA}

\author[0000-0001-7151-009X]{Nikko J. Cleri}
\affiliation{Department of Astronomy \& Astrophysics, The Pennsylvania State University, University Park, PA 16802, USA}
\affiliation{Institute for Computational \& Data Sciences, The Pennsylvania State University, University Park, PA 16802, USA}
\affiliation{Institute for Gravitation and the Cosmos, The Pennsylvania State University, University Park, PA 16802, USA}

\author[0000-0003-3881-1397]{Olivia R. Cooper}\altaffiliation{NSF Graduate Research Fellow}
\affiliation{The University of Texas at Austin, 2515 Speedway Boulevard Stop C1400, Austin, TX 78712, USA}

\author[0000-0001-8460-1564]{Pratika Dayal}
\affiliation{Kapteyn Astronomical Institute, University of Groningen, P.O. Box 800, 9700 AV Groningen, The Netherlands}

\author[0000-0001-7201-5066]{Seiji Fujimoto}\altaffiliation{Hubble Fellow}
\affiliation{Department of Astronomy, The University of Texas at Austin, Austin, TX 78712, USA}
\affiliation{David A. Dunlap Department of Astronomy and Astrophysics, University of Toronto, 50 St. George Street, Toronto, Ontario, M5S 3H4, Canada}

\author[0000-0001-6278-032X]{Lukas J. Furtak}
\affiliation{Department of Physics, Ben-Gurion University of the Negev, P.O. Box 653, Be'er-Sheva 84105, Israel}

\author[0000-0003-4700-663X]{Andy~D.~Goulding}
\affiliation{Department of Astrophysical Sciences, Princeton University, 4 Ivy Lane, Princeton, NJ 08544, USA}

\author[0000-0002-3301-3321]{Michaela Hirschmann}
\affiliation{Institute of Physics, GalSpec laboratory, EPFL, Observatory of Geneva, Chemin Pegasi 51, 1290 Versoix, Switzerland}

\author[0000-0002-5588-9156]{Vasily Kokorev}
\affiliation{Department of Astronomy, The University of Texas at Austin, Austin, TX 78712, USA}

\author[0000-0003-0695-4414]{Michael V.\ Maseda}
\affiliation{Department of Astronomy, University of Wisconsin-Madison, 475 N. Charter St., Madison, WI 53706 USA}

\author[0000-0002-2446-8770]{Ian McConachie}
\affiliation{Department of Astronomy, University of Wisconsin-Madison, 475 N. Charter St., Madison, WI 53706 USA}

\author[0000-0003-2871-127X]{Jorryt Matthee}
\affiliation{Institute of Science and Technology Austria (ISTA), Am Campus 1, 3400 Klosterneuburg, Austria}

\author[0000-0001-8367-6265]{Tim B. Miller}
\affiliation{Center for Interdisciplinary Exploration and Research in Astrophysics (CIERA), Northwestern University,1800 Sherman Ave, Evanston, IL 60201, USA}

\author[0000-0003-3729-1684]{Rohan P. Naidu}\altaffiliation{Hubble Fellow}
\affiliation{$^1$MIT Kavli Institute for Astrophysics and Space Research, 70 Vassar Street, Cambridge, MA 02139, USA}

\author[0000-0001-5851-6649]{Pascal A. Oesch}
\affiliation{Department of Astronomy, University of Geneva, Chemin Pegasi 51, 1290 Versoix, Switzerland}
\affiliation{Cosmic Dawn Center (DAWN), Niels Bohr Institute, University of Copenhagen, Jagtvej 128, K{\o}benhavn N, DK-2200, Denmark}

\author[0000-0002-9651-5716]{Richard Pan}
\affiliation{Department of Physics \& Astronomy, Tufts University, 574 Boston Avenue, Medford, MA 02155, USA}

\author[0000-0002-0108-4176]{Sedona H. Price}
\affiliation{Department of Physics \& Astronomy and PITT PACC, University of Pittsburgh, Pittsburgh, PA 15260, USA}

\author[0000-0002-1714-1905]{Katherine~A.~Suess}
\affiliation{Department for Astrophysical \& Planetary Science, University of Colorado, Boulder, CO 80309, USA}

\author[0000-0003-1614-196X]{John R. Weaver}
\affiliation{Department of Astronomy, University of Massachusetts, Amherst, MA 01003, USA}

\author[0000-0003-1207-5344]{Mengyuan Xiao}
\affiliation{Department of Astronomy, University of Geneva, Chemin Pegasi 51, 1290 Versoix, Switzerland}

\author[0000-0001-6454-1699]{Yunchong Zhang} 
\affiliation{Department of Physics and Astronomy and PITT PACC, University of Pittsburgh, Pittsburgh, PA 15260, USA}

\author[0000-0002-0350-4488]{Adi Zitrin}
\affiliation{Department of Physics, Ben-Gurion University of the Negev, P.O. Box 653, Be'er-Sheva 84105, Israel}

\begin{abstract}

Luminous broad H$\alpha$ emission and red rest-optical SEDs are the hallmark of compact Little Red Dots (LRDs), implying highly attenuated dusty starbursts and/or obscured active galactic nuclei. However, the lack of observed FIR emission has proved difficult to reconcile with the implied attenuated luminosity in these models. Here, we utilize deep new ALMA imaging, new and existing JWST/MIRI imaging, and archival Spitzer/Herschel imaging of two of the rest-optically brightest LRDs ($z=3.1$ and $z=4.47$) to place the strongest constraints on the IR luminosity in LRDs to date. The detections at $\lambda_\mathrm{rest}=1-4 \ \mu$m imply flat slopes in the rest-IR, ruling out a contribution from hot ($T\gtrsim500$ K) dust. Similarly, FIR non-detections rule out any appreciable cold ($T\lesssim75$ K) dust component. Assuming energy balance, these observations are inconsistent with the typical FIR dust emission of dusty starbursts and quasar torii, which usually show a mixture of cold and hot dust. Additionally, our \cii\ non-detections rule out typical dusty starbursts. We compute empirical maximum IR SEDs and find that both LRDs must have $\log(L_\mathrm{IR}/L_\odot) \lesssim 12.2$ at the $3\sigma$ level. These limits are in tension with the predictions of rest-optical spectrophotometric fits, be they galaxy only, AGN only, or composite. It is unlikely that LRDs are highly dust-reddened intrinsically blue sources with a dust temperature distribution that conspires to avoid current observing facilities. Rather, we favor an intrinsically redder LRD SED model that alleviates the need for strong dust attenuation.
\end{abstract}


\section{Introduction} \label{sec:intro}

One of the central mysteries of the early JWST era has been the population of compact, red sources that have been dubbed ``Little Red Dots" \citep[LRDs,][]{Matthee2024}. While a number of LRD selections that identify a range of compact red objects with varying physical properties are employed in the literature \citep[e.g.,][]{Greene2024, Kokorev2024_LRD, Kocevski2024, PerezGonzalez2024_LRD,Barro2024_warmdust}, in this work we are specifically interested in the population of sources with a characteristic ``V shaped" transition between the rest optical and red continuum that tends to be associated with the Balmer limit \citep[e.g.,][]{Wang2024_BRD, Ma2024_LRD, Labbe2024_monster, Setton2024_LRDbreak, Ji2025_LRD}, broad and luminous Balmer emission \citep[which is often attributed to an AGN broad line region, e.g.,][]{Kokorev2023_LRD, Killi2023, Greene2024, Kocevski2024, Matthee2024}, and unresolved rest-optical continuum emission \citep[e.g.,][]{Furtak2023_LRD,Greene2024, Labbe2025_LRD, Matthee2024}.

These sources have a high number density \citep[$\rho \sim 10^{-5}$ cMpc$^{-3}$ at $4<z<8$,][]{Greene2024, Akins2024, Kokorev2024_LRD, Kocevski2024, Matthee2024, Taylor2024}, making understanding their properties crucial to the study of galaxy and AGN evolution. However, there has been no clear consensus on the engine that powers their luminous red continua. Because of the extremely red colors of LRDs, essentially every model of their light invokes a highly reddened galaxy \citep[e.g.,][]{Labbe2023UB,Akins2023, PerezGonzalez2024_LRD, Baggen2023,Baggen2024, Williams2024,Barro2024, Akins2024,Rinaldi2024}, a highly reddened quasar \citep[e.g,][]{Kocevski2024, Greene2024, Li2024_LRD, Labbe2025_LRD, Ji2025_LRD,Barro2024, Akins2024,Rinaldi2024}, or composite models that are a combination of the two components \citep[][]{Wang2024_BRD, Wang2024_UB, Ma2024_LRD, Leung2024, Juodzbalis2024, Barro2024_warmdust}. While the specific ingredients and details of these models vary, they all share the common assumption that the engine of the LRD that ionizes the broad lines is intrinsically blue, and that the red color comes from the attenuation of that engine.

Because highly reddened sources re-emit their attenuated luminosity as dust emission in the infrared (IR), there has been considerable interest in constraining the IR properties of LRDs. However, the observed properties of small samples have in some ways made the LRD SED more confusing. Spatially extended red sources at $2<z<7$ that occupy a similar color space to LRDs are commonly detected at 1.2mm, indicating that cold dust is emitting reprocessed UV radation. In contrast, LRDs are \emph{never} detected at 1.2 mm, even in stacks \citep{Labbe2025_LRD, Williams2024}, casting doubt on the viability of dusty starburst solutions. However, LRDs also lack the characteristic steeply rising hot ($T \gtrsim 500$ K) dust emission at $\lambda_\mathrm{rest}\sim1-5 \ \mu$m commonly seen in AGN, whether in individual luminous sources \citep{Wang2024_BRD, Barro2024_warmdust} or in stacks \citep{Williams2024}, meaning that if they are indeed reddened quasars, they do not present similarly in the IR to dusty AGN seen locally \citep[e.g.,][]{Polletta2007} or at cosmic noon \citep[e.g.,][]{Glikman2012,Assef2016, Hamann2017, Ma2024_ERQ}.

The most extensive stacking analysis of LRDs to date was carried out in \cite{Akins2024}, leveraging the wide area of COSMOS-Web \citep{Casey2023_CW} to photometrically select hundreds of LRDs at $z=5-9$, many of which were covered by MIRI imaging and legacy mid- and far-IR observations. Their stack consists entirely of non-detections at $\lambda_\mathrm{rest}>2 \ \mu$m, but it represents the best limits on the maximal IR SED of a typical LRD. \cite{Li2024_LRD} use this stack to demonstrate that a physically extended dust distribution that outputs the majority of its luminosity at $\lambda_\mathrm{rest} = 10-100 \ \mu$m could be consistent with energy balance solutions of reddened quasars. Additionally, \cite{Casey2024} argue that the compact sizes (and consequently, their low dust masses) should indeed result in dust temperatures that peak in that same wavelength range, with IR SEDs that remain undetectable by sub-mm facilities like ALMA. While these works represent an important step forward in constraining the dust SED of LRDs, they are subject to all the usual limitations of stacked analyses of heterogeneous samples, especially given that the \cite{Akins2024} LRDs were selected only on red color and compactness, not ``V shape" or the explicit detection of a broad line. Moreover, because the typical source included in the stack is optically faint, the constraints on $L_\mathrm{IR}/L_\mathrm{optical}$ are still quite weak for the ensemble population.

However, while the kinds of shallow mid- and far-IR observations of typical LRDs that have been achieved in un-targeted wide field observations are not significantly constraining for energy balance, deep observations of the most luminous individual LRDs still hold promise for constraining the dust luminosity and temperature. 
In this work, we present new and existing deep MIRI and new ALMA observations of two sources: UNCOVER-45924 \citep{Labbe2024_monster} and RUBIES-BLAGN-1 \citep{Wang2024_BRD}, two of the most luminous LRDs identified to date, with $\nu L_\nu$ at rest-frame 1 $\mu$m a factor of 10 brighter than the typical source included in the \cite{Akins2024} stacks. Both targets have extensive NIRCam photometric coverage, low- and medium-resolution spectroscopy of their continuum and line emission, and archival Spitzer and Herschel imaging \citep[][S. McNulty et al. in preparation]{Dickinson2003,Egami2010}, marking them as the best candidates identified to date for this kind of targeted follow-up.

This letter is laid out as follows. In Section \ref{sec:data}, we describe the two sources studied in this work and present our new observations. In Section \ref{sec:analysis}, we compare the IR luminosity predictions based on the SED fits presented in \cite{Labbe2024_monster} and \cite{Wang2024_BRD} to our new IR constraints, using established dust templates as well as simple blackbody descriptions of the dust SED. Finally, in Section \ref{sec:discussion}, we discuss the implications of these constraints for our understanding of the physical properties of LRDs. Throughout this work, we adopt the best-fit cosmological parameters from the WMAP 9 year results \citep{Hinshaw2013}: $H_0 = 69.32 \ \mathrm{km \ s^{-1} \ Mpc^{-1}}$, $\Omega_m = 0.2865$, and $\Omega_\Lambda = 0.7135$, utilize a Chabrier initial mass function \citep{Chabrier2003}, and quote AB magnitudes.

\section{Data} \label{sec:data}

\subsection{The two most H$\alpha$ luminous LRDs in the southern sky}

At the time of writing, there does not exist a study of LRDs that have spectroscopically confirmed broad lines, high signal-to-noise PRISM spectroscopy (spanning the characteristic ``V shape"), MIRI imaging (constraining the presence of hot dust), \emph{and} deep ALMA observations (constraining cold dust). The goal of this work is to rectify that by unifying such observations for two of the most luminous LRDs in the southern sky where such panchromatic constraints are obtainable. To accomplish this, in this work we focus our study on A2744-45924 \citep{Labbe2023UB} and RUBIES-BLAGN-1 \citep{Wang2024_BRD}, which host highly luminous H$\alpha$ ($\sim10^{43.5}$ erg/s, before applying any dust correction) and are bright in the rest optical ($m_{F444W}\sim22$). The combination of these features makes them the best targets for obtaining solid constraints on the mid- and far-IR properties of this puzzling population.

At $z=4.4655$ and with $\mu=1.7$ \citep{Furtak2023_lensing}, A2744-45924 is the brightest LRD in the A2744 field. The source was first spectroscopically confirmed to have a broad line in \cite{Greene2024}, and full spectral modeling of the source with a wide array of AGN and stellar models was presented in \cite{Labbe2024_monster}. As a part of the UNCOVER \citep[JWST-GO \#2561, PIs: Labbe and Bezanson,][]{Bezanson2022b} and MegaScience \citep[JWST-GO \#4111, PI: Suess,][]{Suess2024} programs, the source has been observed with all NIRCam broad and medium bands \citep[catalogued in][]{Weaver2024}, as well as with deep NIRSpec PRISM spectroscopy \citep[$\sim$16.3 hours][]{Price2024} and NIRCam/GRISM F365W spectroscopy that observes H$\alpha$ \citep[ALT, JWST-GO \#3516,][]{Naidu2024_alt}. It is unresolved in the rest-optical, with a magnification corrected size $r_e<70$ pc in F200W, though at short wavelengths there is a secondary component seen in the bluest NIRCam bands that is, if it is at that same redshift, separated by $\sim700$ pc \citep{Labbe2024_monster}. It is host to an extremely broad (FWHM=$4500$ km/s) H$\alpha$ component, and, additionally, both redshifted and blueshifted narrow absorption is detected at $\Delta v\sim150$ km/s relative to the systemic velocity \citep{Labbe2024_monster}. In addition to its highly luminous H$\alpha$, A2744-45924 also hosts a wide range of extremely high-equivalent width UV emission lines \citep{Treiber2024, Labbe2024_monster}.

At $z=3.1$, RUBIES-BLAGN-1 is the most luminous LRD in the UDS field, where it was observed in 8 NIRCam Bands (F090W, F115W, F150W, F200W, F277W, F356W,
F444W and F410M) as well as two MIRI bands (F770W and F1800W) as part of the PRIMER Survey \citep{Donnan2024}. NIRSpec/PRISM and NIRSpec/G395M spectroscopy was obtained as part of the RUBIES program \citep[JWST-GO \#4233, PIs: PIs de Graaff and Brammer;][]{deGraaff2023}, and was presented along with full spectrophotometric modeling in \cite{Wang2024_BRD}. Even at PRISM resolution, the H$\alpha$ emission is clearly broad, and the source also exhibits broad (FWHM$\sim4000$ km/s) Pa$\delta$, Pa$\gamma$, HeI emission, with evidence for blueshifted HeI absorption at $\Delta v = -210$ km/s \citep{Wang2024_BRD}. The source was the first individual spectroscopically confirmed LRD to be detected in MIRI, with a flat rest-frame $1-4 \ \mu$m color that showed no indication of a significant luminosity contribution from hot torus dust.

\begin{figure*}
    \centering
    \includegraphics[width=\textwidth]{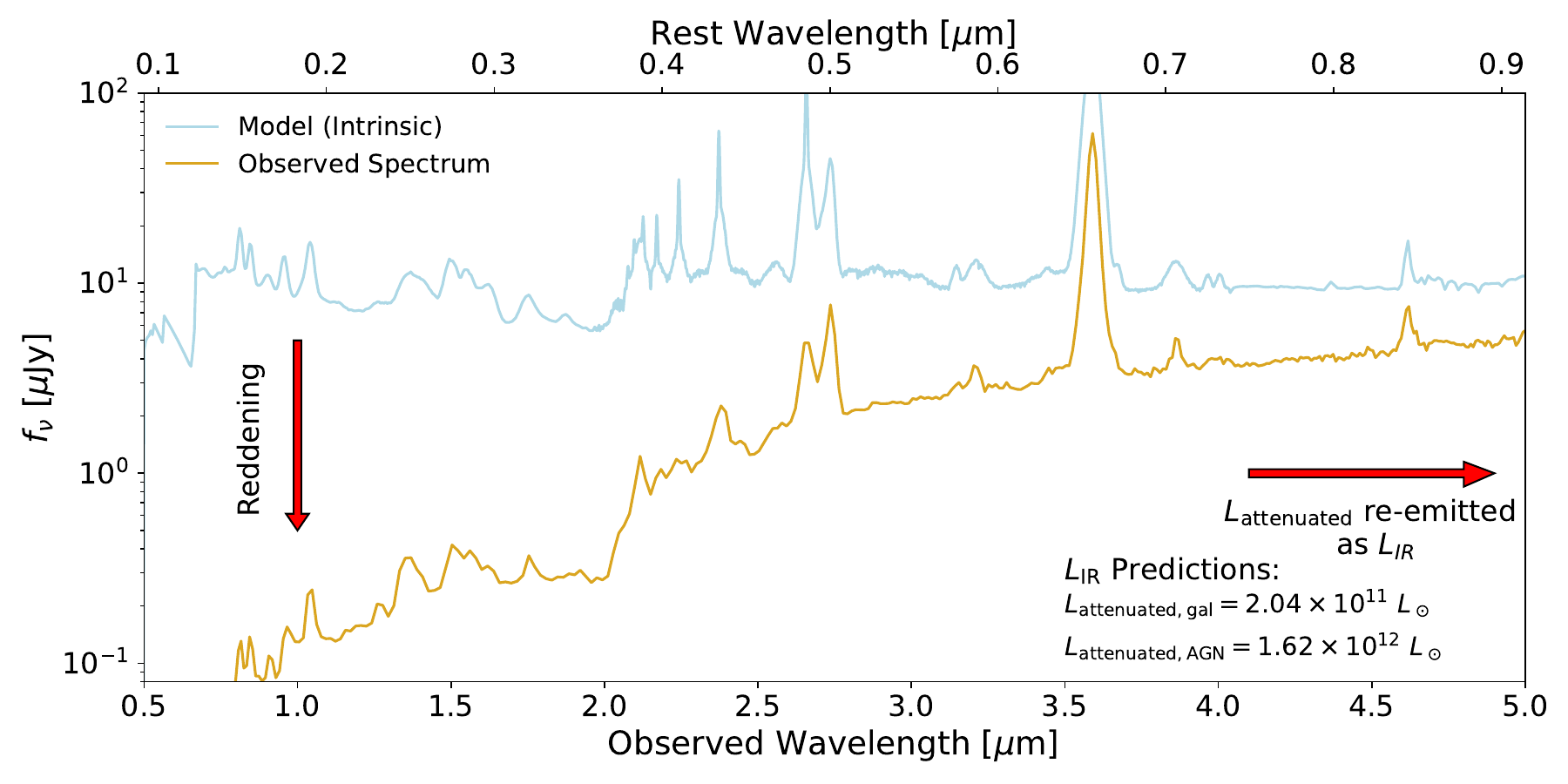}
    \caption{A demonstration of the fundamental assumption that goes into the vast majority of LRD models, using the composite model from \cite{Labbe2024_monster} for A2744-45924. In blue, we show the intrinsic, unattenuated galaxy+AGN model, which, via a combination of reddening and scattering, is observed in the rest optical as the NIRSpec/PRISM spectrum (orange). The total attenuated luminosity for both the galaxy and AGN components are listed in the bottom right. Assuming energy balance, this model, and models like it, predict significant FIR output where the dust that is being heated by the intrinsically blue engine of the LRD, regardless of whether that engine is primarily driven by a starburst or an AGN.}
    \label{fig:monster_atten_demo}
\end{figure*}

\subsection{Restframe-optical SED modeling} \label{subsec:models}

Previous works have modeled a number of high SNR LRD UV/optical spectra with complementary methods \citep{Ma2024_LRD, Wang2024_BRD, Wang2024_UB, Labbe2024_monster}. They have used a range of star formation history parameterizations to describe the possible contribution from stars and a number of different power-law prescriptions to describe a contribution from a standard AGN disk. In all cases, it is possible to derive acceptable fits with a wide range of mixtures between AGN and galaxy contribution, provided that all UV-luminous components (be they AGN and star formation) are significantly reddened by a very steep dust law. Producing the observed red continuum and sharp transition of the ``V shape" at the Balmer limit \emph{always} requires a significantly reddened component that is intrinsically UV-luminous. This reflects the vast majority of the work in the literature, which also reaches conclusions that either highly luminous dust obscured AGN and/or starbursts are powering LRDs.

In this work, we directly utilize the best fitting models for A2744-45924 and RUBIES-BLAGN-1 from \cite{Labbe2024_monster} and \cite{Wang2024_BRD}. We refer the reader to these works for the specific ingredients of each of these models, but in practice, they function similarly by fitting galaxy-only, AGN-only, or composite models to the observed spectra. In particular, we utilize the predictions from these models for the total attenuated luminosity, which should, under the assumption of energy balance, be equal to the total IR luminosity. Despite the range of physical ingredients for these models and the real differences in the rest-optical spectral shape of the sources, these models require a similar range of attenuated luminosity to produce the sharp ``V shape" and red continuum, predicting highly luminous IR SEDs with $\log(L_{IR}/L_\odot)=12.3-12.8$. We note that \cite{Barro2024_warmdust} also performed SED fitting analysis for RUBIES-BLAGN-1 and arrived at a significantly lower IR luminosity due to an intrinsically fainter AGN. However, the modeling performed by \cite{Barro2024_warmdust} was based only on photometry and failed to capture the Balmer break that is apparent in the observed PRISM spectrum. Thus, we proceed with our analysis using the higher LIR values that come from the spectroscopic modeling in \cite{Wang2024_BRD}.

To demonstrate the level of dust-attenuation that these models require ($A_V=2.4-3$ mag in all models presented in \citealt{Wang2024_BRD} and \citealt{Labbe2024_monster}), in Figure \ref{fig:monster_atten_demo} we show the de-reddened best fitting composite AGN+galaxy model (blue) from \cite{Labbe2024_monster} in addition to the observed spectrum (goldenrod). The best-fitting galaxy+AGN model predicts a total attenuated luminosity of $\sim1.8\times10^{12}L_\odot$, with $\sim90\%$ of that luminosity produced by the AGN component and the remaining 10\% coming from a reddened post-starburst galaxy that produces the break and contributes to the red continuum. 

Assuming these two LRDs are drawn from random viewing angles such that $\langle \frac{L_\mathrm{attenuated}}{L_\mathrm{IR}} \rangle \sim 1$, this entire budget of attenuated luminosity should be re-emitted by continuum dust emission that peaks somewhere in the mid/far-IR, depending on the dust temperature. In the next section, we utilize this as motivation for our follow-up observations to constrain this predicted dust distribution across a wide range of assumed temperatures.

\subsection{IR Constraints} \label{subsec:IR_data}

Given that models make clear predictions for the IR luminosity but are agnostic to the specific dust temperature or composition, we conduct a wide search for reprocessed dust emission across wavelengths that are accessible to our current observing facilities. In this section, we outline our motivation for each IR data source, as well as our observing strategy in reductions when new data was taken. All the measurements from this subsection, along with published measurements of interest, are presented in Table \ref{tbl:data}.

\begin{deluxetable}{ccc} \label{tbl:data}

\tablehead{
\colhead{} & \colhead{\textbf{A2744-45924}} & \colhead{\textbf{RUBIES-BLAGN-1}}
}
\tablecaption{Observed properties of the two LRDs studied in this work. All fluxes are in observed units and have not been corrected for the effects of lensing, but all luminosities have been. No measurements were corrected for any assumed dust attenuation. All fluxes are in units of $\mu$Jy, and all luminosities are in units of $L_\odot$. All upper limits are quoted at the $3\sigma$ level.}
\startdata
$z_{spec}$ & 4.4655  & 3.1034 \\
$\mu$ & $1.7\pm0.2$ &  1 \\
\hline
Flux Density [$\mu$Jy]: & & \\
MIRI/F770W & -- & $8.9\pm0.4$ \\
MIRI/F1000W & $8.6\pm0.3$ & -- \\
MIRI/F1800W & -- & $13.0\pm0.6$ \\
MIRI/F2100W & $9.0\pm0.9$ & -- \\
Spitzer/MIPS 24$\mu$m & -- & $<30$ \\
Herschel/PACS 100 & $<1700$ & $<1500$ \\
Herschel/PACS 160 & $<9200$ & $<3300$ \\
ALMA/Band 9 & $<435$ & -- \\
ALMA/Band 8 & -- & $<132$ \\
ALMA/Band 7 & $<44$ & -- \\
ALMA/Band 6 & $<125$ & $<21$ \\
\hline
Luminosity [$L_\odot$]: & & \\
$L_{[CII]}$ & $<2.7\times10^7$ & $<6.7\times10^7$ \\
$L_{\mathrm{H}\alpha}$ & $16.9 \pm 0.9 \times10^9$ & $8.9 \pm 0.3 \times10^9$
\enddata
\end{deluxetable}

\subsubsection{MIRI}

Given that a reddened AGN is commonly invoked to model the red continuum of LRDs \citep[e.g.,][]{Onoue2023,Labbe2025_LRD,Furtak2024,Wang2024_BRD, Wang2024_UB, Ma2024_LRD}, it stands to reason that the host dust torii near the black hole that emit at $\lambda_\mathrm{rest} \gtrsim 1 \ \mu$m typically observed in reddened quasars \citep[e.g.,][]{Assef2016, Hamann2017, Ma2024_ERQ, Bosman2024} would be seen in LRDs. However, to date, rest-MIR detctions have been consistent with the Rayleigh-Jeans tail of an accretion disk without the need for an additional hot dust component, even in stacks \citep{Williams2024, Akins2024}. RUBIES-BLAGN-1 already has mid-IR imaging in the MIRI F770W and F1800W filters ($\lambda_\mathrm{rest}\sim1.5, 3.5 \ \mu$m) from the PRIMER survey (JWST-GO-1837), exhibiting no evidence for a hot dust torus \citep{Wang2024_BRD}. For our analysis, we utilize the fluxes reported in that work, $8.9\pm0.4$ and $13.0\pm0.6$ $\mu$Jy in F770W and F1800W, respectively.

A2744-45924 was observed in MIRI/F1000W and MIRI/F2100W on October 30, 2024 for 11 minutes in F1000W and 30 minutes in F2100W as a part of JWST-GO \#6761 (PI: Greene), with depths that were chosen to be able to detect a flat continuum similar to what is seen in RUBIES-BLAGN-1 and other LRDs \citep[including those well-fit with stellar continuum from a post-starburst population alone, see][]{Williams2024}. Our observations utilized the FASTR1 readout mode and 4 dithers, with 60 groups and 150 groups per exposure in F1000W and F2100W filters, respectively, following the SMILES survey strategy \citep{Alberts2024}.  Data reduction was performed using the JWST calibration pipeline v1.12.5 with custom steps for warm pixel removal and iterative background subtraction as described in \citet{Alberts2024}, as well as the masking of partially saturated pixels across subsequent dithers to remove persistence artifacts (S. Alberts et al. 2025, in prep).  Astrometry for the F1000W image was corrected using the UNCOVER photometric catalog and an astrometry-corrected F1000W catalog was then used to correct the F2100W image, to ensure the maximum number of matched high-SNR detections.

Following \cite{Alberts2024}, we measure the fluxes in F1000W and F2100W using aperture sizes that enclose 65\% of the energy of a point source ($r=$ 0.36'' and 0.6'', respectively), which we correct to total under the assumption that our sources are unresolved. Rather than using the pipeline uncertainty vector, we assume that the uncertainty in each pixel is equal to the standard deviation of the full image. The measured fluxes are $8.6\pm0.3$ and $9.0\pm0.9$ $\mu$Jy in F1000W and F2100W, respectively.

\begin{figure*}
    \centering
    \includegraphics[width=\textwidth]{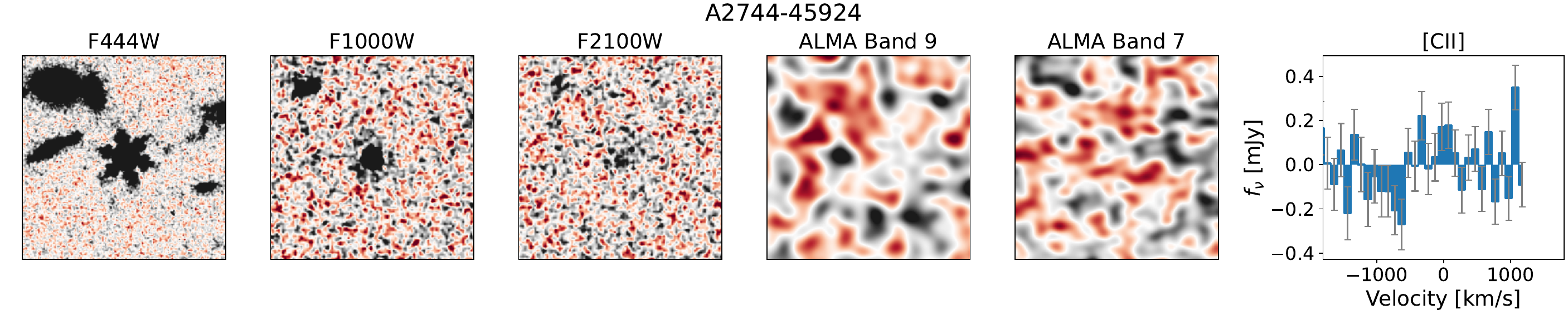}
    \includegraphics[width=\textwidth]{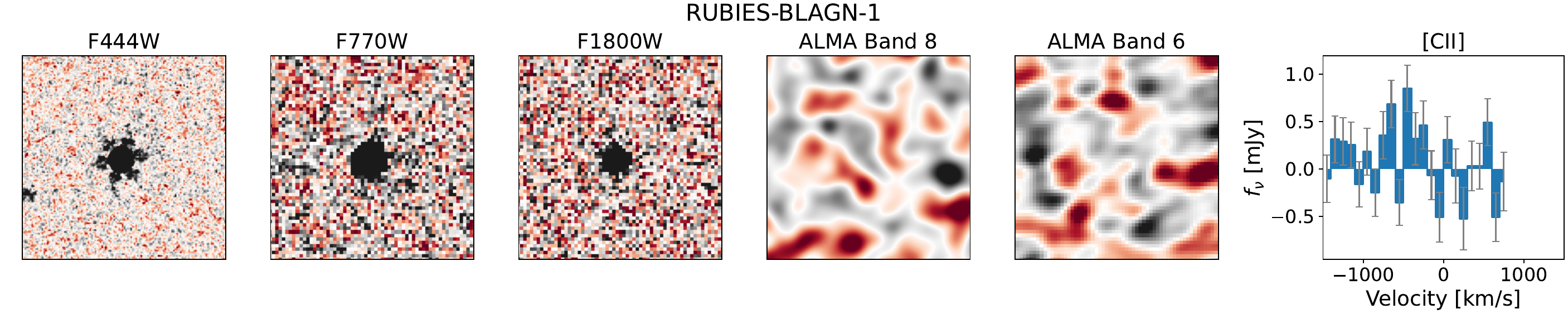}
    \caption{6"x6" cutouts of A2744-45924 (top) and RUBIES-BLAGN-1 (bottom), showcasing the F444W imaging \citep[A2744-45924:][RUBIES-BLAGN-1: JWST-GO-1837; PI Dunlop]{Bezanson2022b, Weaver2024}, as well as MIRI imaging (this program for A2744-45924, PRIMER for BRD-BLAGN-1), and new FIR continuum imaging and spectroscopy obtained with ALMA. Both sources are detected in the mid-IR but not in the far-IR. Additionally, neither source is significantly detected in \cii.}
    \label{fig:cutouts}
\end{figure*}

\subsubsection{ALMA}

For both target LRDs, we carried out new ALMA observations in two bands in project 2024.1.00826.S (PI: J. Greene). One band targets the \cii 158\,\um line and underlying rest-frame $\approx$160\,\um continuum emission, and the other band targets the dust continuum emission at an observed frequency that depends on the redshift of each source. For A2744-45924, \cii falls in the ALMA Band~7 receiver coverage at 347.82\,GHz, and we also observed the continuum at $\sim$690\,GHz in Band~9. For RUBIES-BLAGN-1, \cii falls in ALMA Band~8 at 463.15\,GHz, and we additionally observed the continuum at $\sim$240\,GHz in Band~6. For the \cii observations, the correlator was configured to provide $\approx$3.74\,GHz of contiguous bandwidth around the \cii sky frequency and 7.81\,MHz channels; the alternate sideband used 31.25\,MHz channels. The continuum-only observations of each target used the observatory-defined standard Band~9 or 6 continuum configurations.

Band~7 observations of A2744-45924 were executed on 2024 October 14 and 15 for a total on-source time of 97\,min. The array consisted of 47 or 48 antennas on baselines spanning 15--500\,m, producing a $\approx$0.7'' synthesized beam with natural visibility weighting. The continuum sensitivity with this weighting is 14.4\,\uJy. We also produced \cii cubes, which reach a sensitivity of 110\,\uJy in 100\,\kms channels. The atmospheric transmission is smooth at the \cii frequency, and we verify that the quoted depth scales as expected for narrower or wider velocity channels. The Band~9 observations were carried out on 2024 October 13 for a total on-source time of 99\,min in excellent weather conditions. With natural weighting, the synthesized beam size is $\approx$0.3'' with 110\,\uJy sensitivity. In order to mitigate concerns that this high resolution may resolve out any extended host galaxy emission, we also apply a 0.4'' Gaussian uv taper; the resulting image has a $\approx$0.55'' synthesized beam and 145\,\uJy sensitivity.

RUBIES-BLAGN-1 was observed in Band~8 on 2024 October 3 and 12 for a total of 198\,min on-source with 47 and 44 antennas, respectively, and baselines ranging from 15--500\,m. The synthesized beam with natural weighting is $\sim$0.45'', but we again apply a uv taper to avoid resolving the target galaxy. The sensitivities below use a 0.5'' taper to reach $\sim$0.75'' angular resolution. The atmospheric transmission is more challenging at the observed Band~8 frequencies, with two main consequences. First, we discard half of the continuum bandwidth in the upper sideband due to its proximity to a deep telluric oxygen feature (placing the continuum coverage in the lower sideband would have faced the same issue, but with the 448\,GHz water line). The continuum sensitivity in the tapered image is 54\,\uJy. Second, the \cii sky frequency is close to a narrow ozone line, resulting in $\approx$35\% worse sensitivity in a 100\,\kms bandwidth centered at $-$30\,\kms. The consequence is that wider velocity channels include more data with better transmission; the naturally-weighted cubes reach sensitivities of 330, 170, and 125\,\uJy for 100, 300, and 500\,\kms channels, respectively, at the expected \cii frequency. Band~6 observations were carried out on 2024 October 18 and 19 for 115\,min on-source. With natural visibility weighting, the synthesized beam is $\approx$1.0'' and the data reach 7.0\,\uJy sensitivity. 

We do not detect continuum or \cii emission from either source. We place upper limits on the continuum flux density using the rms of each image (tapered, as noted above); this assumes the sources would be pointlike at the resolution of the data. We extract \cii spectra under the same assumption (from the tapered cube, in the case of RUBIES-BLAGN-1) by fitting for the amplitude of a 2D Gaussian profile with the dimensions of the synthesized beam and centroid fixed to the phase center in each channel of the cubes. We place upper limits on the integrated \cii line luminosity assuming a line width FWHM of 500\,\kms.

\subsubsection{Ancillary FIR Data}

As an additional constraint on the presence of hot dust in RUBIES-BLAGN-1, we utilize existing deep Spitzer/MIPS 24$\mu$m imaging \citep{Dickinson2003} that does not detect the source \citep{Wang2024_BRD, Barro2024_warmdust}. As this source is in 3D-HST catalogs, we utilize the flux and uncertainty reported in \cite{Whitaker2014} to arrive at a $3\sigma$ upper limit of 30 $\mu$Jy. To constrain dust at rest-frame $\sim20-40 \mu$m, we utilize existing Herschel/PACS 100 and 160 $\mu$m imaging of our sources, which, in contrast with typical LRDs, are so luminous that these limits are relevant. For RUBIES-BLAGN-1, we adopt the limits from the 3D-Herschel project (S. McNulty et al. in preparation, NASA-ADAP-80NSSC20K0416). For A2744-45924, we utilize imaging from the Herschel Lensing Survey \citep{Egami2010}. We obtain reduced imaging products from the Herschel Science Archive and perform a 2D sky subtraction with SEP \citep{Bertin1996, Barbary2016}. We measure the flux and uncertainty in 4 and 6 arcsecond apertures at the source location and measure the total flux by multiplying by 2.5 and using the local RMS to estimate uncertainty. We do not detect either image, and our 3$\sigma$ upper limits are 1.7 mJy and 9.2 mJy at 100 and 160 $\mu$m respectively.

\begin{figure*}
    \centering
    \includegraphics[width=0.85\textwidth]{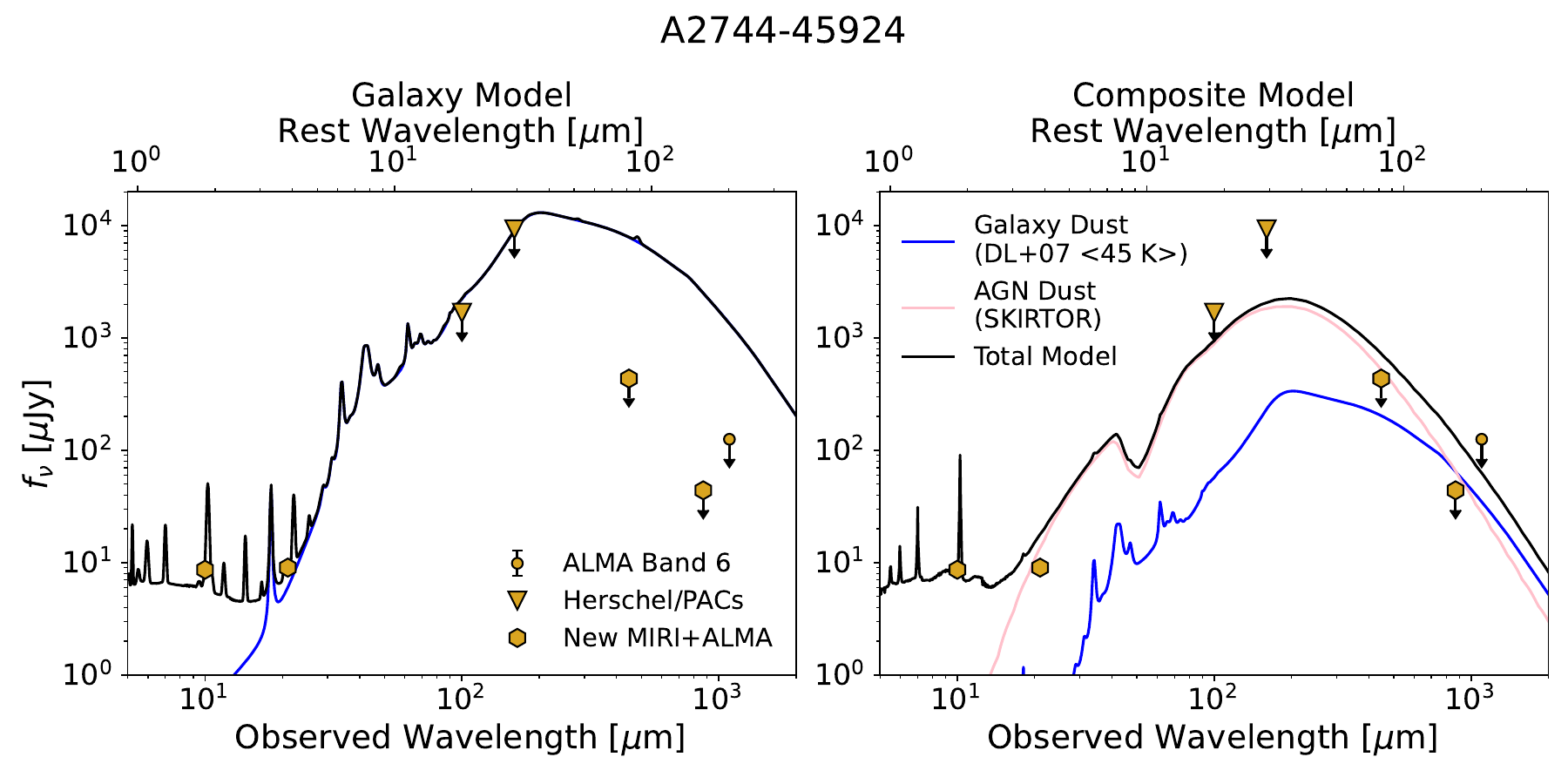}
    \includegraphics[width=0.85\textwidth]{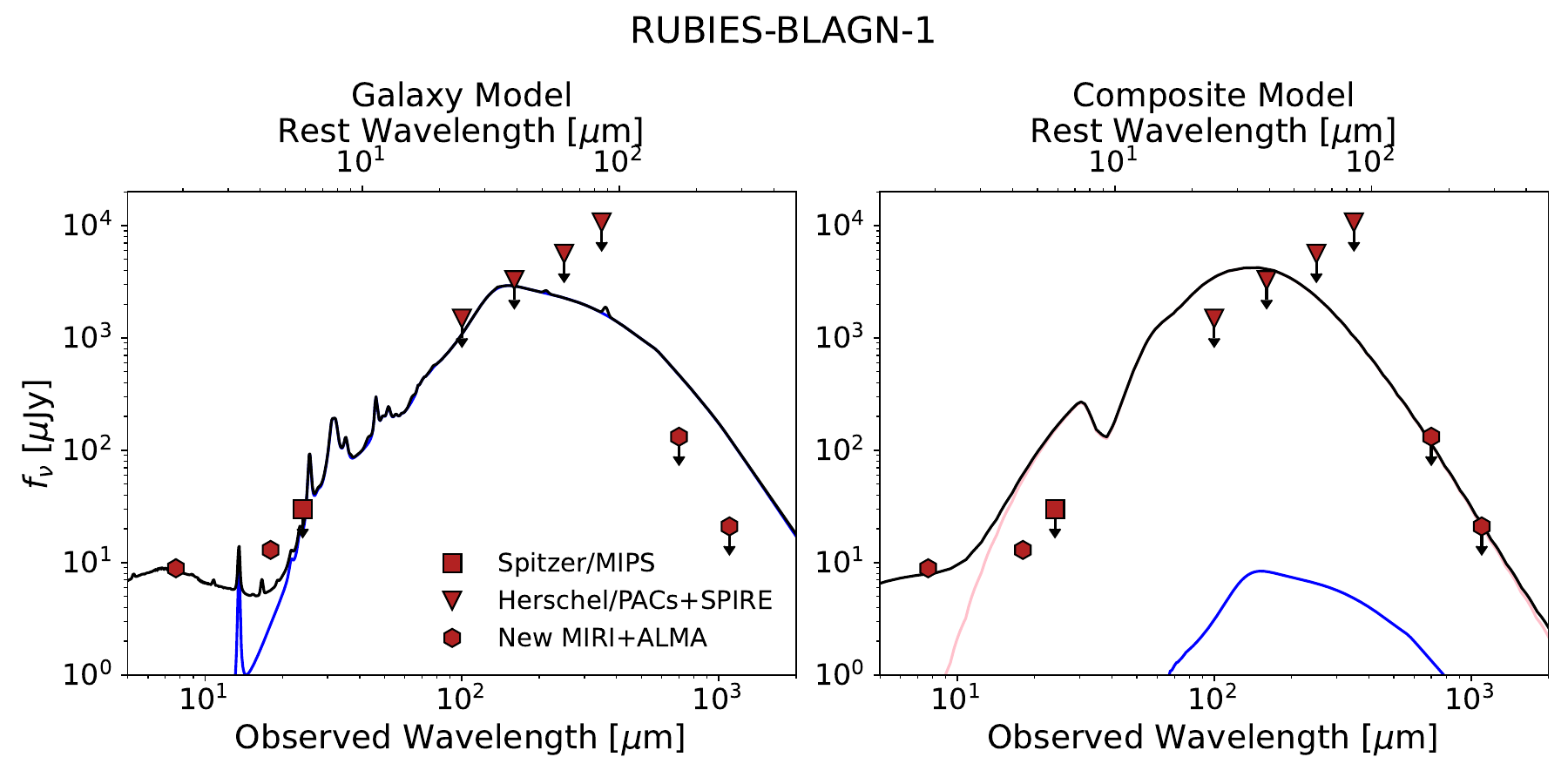}
    \caption{The IR spectral energy distributions of A2744-45924 (top) and RUBIES-BLAGN-1 (bottom). Herschel PACs and SPIRE upper limits are shown as triangles and the Spitzer/MIPS 24$\mu$m for RUBIES-BLAGN-1 is shown as a square. New MIRI and ALMA limits are shown as hexagons. Existing ALMA Band 6 data, in the case of A2744-45924, is shown as a circle. In the first row, we show the predictions for the IR SED using the galaxy-only fits from \cite{Labbe2024_monster} and \cite{Wang2024_BRD}, assuming the dust emits as a $\langle T_\mathrm{dust} \rangle$ = 45 K \cite{Draine2007} template scaled to the attenuated luminosity (blue). On the right, we show the composite models from the same work, with the attenuated luminosity of the galaxy component emitted in the same \cite{Draine2007} template and the attenuated AGN emitted with the coldest SKIRTOR model \citep[][pink]{Stalevski2012, Stalevski2016}. In all cases, the total model is shown in black and is inconsistent with our new constraints.}
    \label{fig:dust_SEDs}
\end{figure*}

\section{Analysis} \label{sec:analysis}

\subsection{Are LRDs consistent with standard galaxy and AGN dust emission SEDs?}

\begin{deluxetable}{cccc} \label{tbl:LIR}
\tablehead{
\colhead{$L_\mathrm{attenuated}$ [$L_\odot$]:} & \colhead{AGN} & \colhead{Galaxy} & \colhead{Total}
}
\tablecaption{Predicted IR luminosities inferred from the best fitting models in \cite{Labbe2024_monster} (A2744-45924) and \cite{Wang2024_BRD} (RUBIES-BLAGN-1). All values for A2744-45924 are corrected for the effect of lensing.}
\startdata
\textbf{A2744-45924} & & & \\
AGN Only & $3.6 \times 10^{12}$ & -- & -- \\
Galaxy Only & -- & $7.9 \times 10^{12}$ & -- \\
Composite & $1.6 \times 10^{12}$ & $2.0 \times 10^{11}$ & $1.8 \times 10^{12}$ \\
\hline
\textbf{RUBIES-BLAGN-1} & & & \\
Galaxy Only & -- & $1.7 \times 10^{12}$ & -- \\
Composite & $3.4 \times 10^{12}$ & $4.8 \times 10^{9}$ & $3.4 \times 10^{12}$ \\
\enddata
\end{deluxetable}

The luminous LRDs we study in this work are characterized by two distinct features in their rest-frame mid- and far-IR SEDs: they exhibit fairly flat colors from $\lambda_\mathrm{rest} = 1-4$ $\mu$m and they are undetected in deep continuum imaging at $\lambda_\mathrm{rest} = 80-200$ $\mu$m. The weak Herschel limits at $\lambda_\mathrm{rest} = \sim20$ $\mu$m also constrain the maximum IR luminosity output in between these tighter JWST and ALMA limits. Armed with these observations and the predictions for the attenuated luminosity from the models described in Section \ref{subsec:models}, in this section we ask whether the predicted IR luminosity can be consistent with observations if the AGN and galaxy dust is emitted in standard SED models that can describe the dust in such systems.

Dust in star forming galaxies is generally dominated in mass by cold grains that emit at $T=10-50$ K, with a hotter component that includes significant emission from polycyclic aromatic hydrocarbons \citep[PAHs, e.g.,][]{Draine2007}. The dust distributions in AGN, in contrast, are thought to peak at much shorter wavelengths due to large contributions for dust near the AGN that is exposed to very hard radiation and heated to $T\sim1000$ K, with additional contributions from warm dust \citep[e.g.,][]{Nenkova2008, Stalevski2012}. 

Here, we set out to test whether such standard dust temperature distributions, as implemented in templates widely used in galaxy and AGN SED fitting, can be consistent with our tight constraints on LRD dust SED. For each model fit to A2744-45924 and RUBIES-BLAGN-1 in \cite{Labbe2024_monster} and \cite{Wang2024_BRD}, respectively, we compute the attenuated luminosity of starlight and AGN separately. These values are shown in Table \ref{tbl:LIR}. We then assume that the dust from the galaxy and AGN are emitted in standard dust templates that we choose to maximize their chance of adhering to the data. The ALMA non-detections imply that there cannot be significant cold dust, so we employ a warm \cite{Draine2007} template with $U_{min}=25$, $\gamma_e=1$, and $q_\mathrm{PAH}=1$ (corresponding to an average dust temperature of 45 K) to represent galaxy dust. Similarly, the flat MIRI colors imply that any contribution from hot dust must be minimal, so to represent AGN dust, we adopt the coldest SKIRTOR \citep{Stalevski2012, Stalevski2016} torus model available in their library, selecting the model that has the maximum fractional energy output at $\lambda_\mathrm{rest}>5 \ \mu$m.

In Figure \ref{fig:dust_SEDs}, we overplot these predicted dust SEDs on the IR SEDs of A2744-45924 (top) and RUBIES-BLAGN-1 (bottom) for the galaxy-only model (left) and the composite galaxy+AGN model (right). We show the total predicted SED in black, and the scaled template galaxy dust and AGN dust SEDs in blue and pink, respectively. Under the above assumptions, it is clear that all of these standard models violate some of our constraints when scaled with the IR luminosity predictions from the rest-optical SED modeling. The warm \cite{Draine2007} model is able to produce the flat mid-IR colors, but vastly over-predicts not only the ALMA far-IR limits, but also the Herschel non-detections, and reductions on the order of 0.5-1 dex in the IR luminosity predicted by the galaxy-only models would be necessary to bring these SEDs into agreement with the far-IR limits. In both composite models, the attenuated AGN dominates the reprocessed luminosity budget, and in both cases, the coldest SKIRTOR SED still predicts too much hot dust \emph{and} too much cold dust to be consistent with our limits, and similar reductions of $>0.5$ dex in the IR luminosity would be necessary to bring these models under all our limits.

Simply put, these standard dust SEDs emit at too wide a range in dust temperatures to adhere to the limits imposed by our deep data at the IR luminosities that are predicted by the standard assumptions of models used in the vast majority of the literature to describe LRDs. While the models fit in \cite{Labbe2024_monster} and \cite{Wang2024_BRD} did not explicitly include these limits, model parameters (e.g., star formation rate, stellar mass, AGN bolometric luminosity) would have to change dramatically to fit the LRD SED without predicting this much IR luminosity. However, due to their compact size, there is considerable reason to believe that the LRD dust SED should not resemble the standard dust SEDs employed here \citep{Casey2024, Li2024_LRD}. As such, in the following section, we explore whether \emph{any} non-parametric dust SEDs constructed from a series of blackbodies can be consistent with the full suite of IR observations in this work under the assumption of energy balance.

\subsection{Empirical Constraints on the IR Luminosity and Dust Temperature} \label{subsec:empirical_dust}

\begin{figure*}
    \centering
    \includegraphics[width=\textwidth]{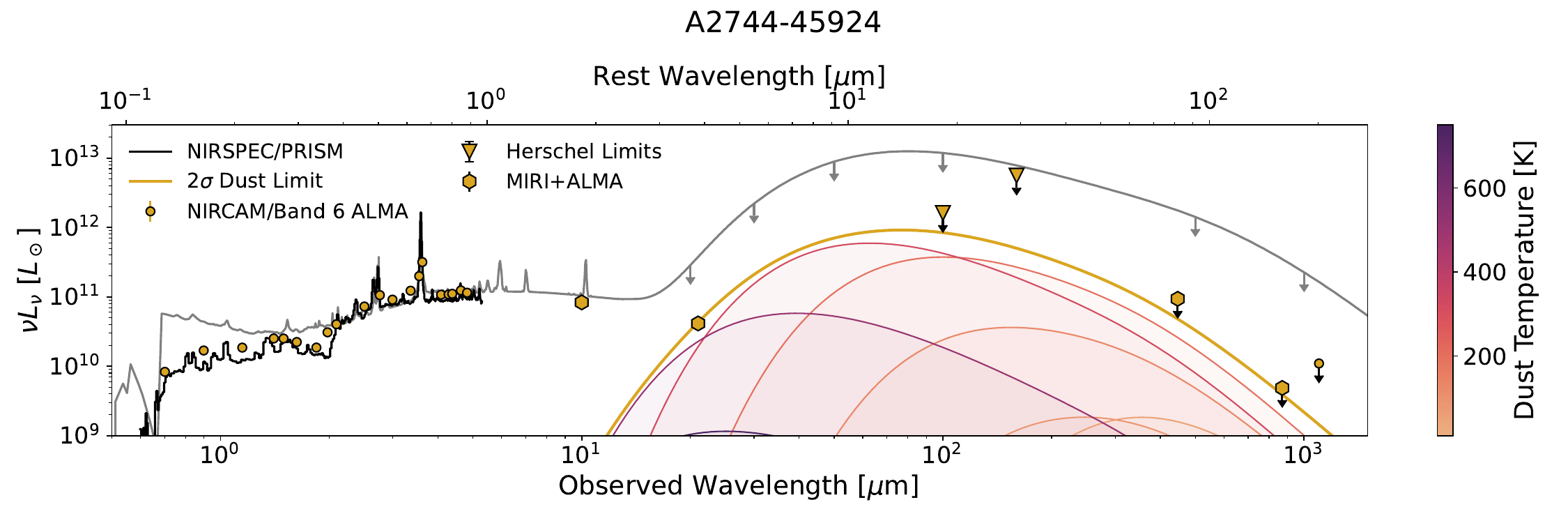}
    \includegraphics[width=\textwidth]{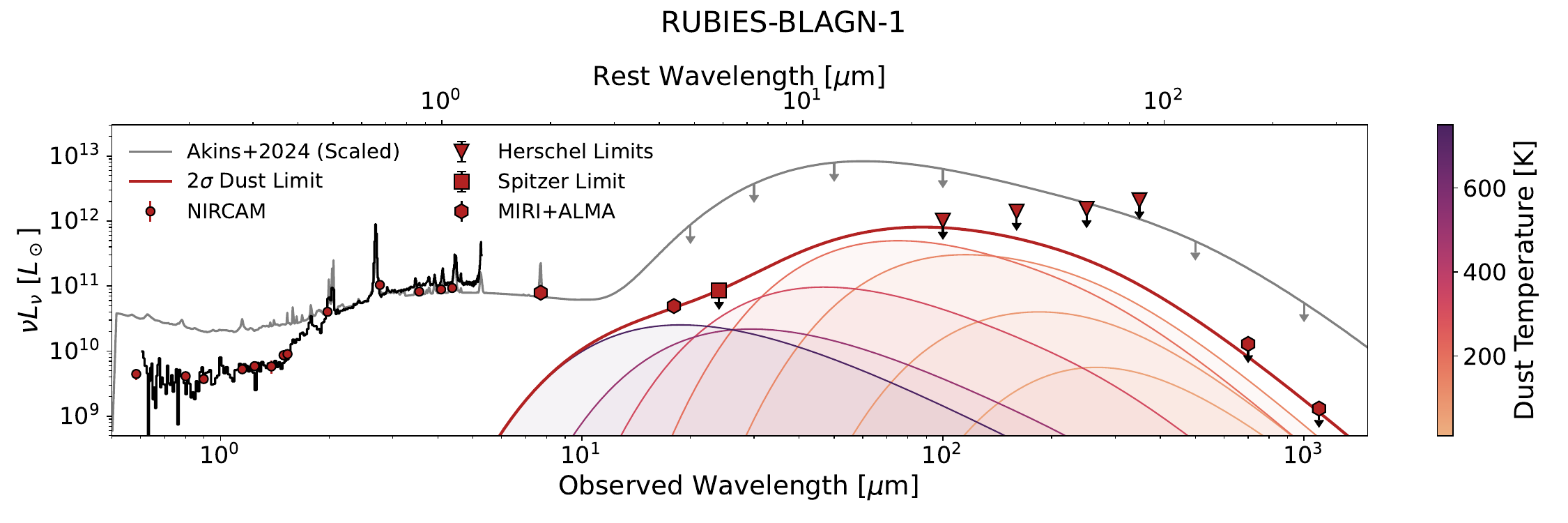}

    \caption{The full SEDs for A2744-45924 (top) and RUBIES-BLAGN-1 (bottom) containing all existing optical-FIR data, with existing NIRCAM photometry shown as circles and NIRCam spectra shown in black. Herschel PACs and SPIRE upper limits are shown as triangles. New MIRI and ALMA limits are shown as hexagons. The colored curves indicate the maximum contribution of single temperature modified $T_\mathrm{dust}$= 20, 30, 50, 130, 200, 320, 510, and 800 K blackbodies to the 3$\sigma$ upper limit FIR SED. The non-detection of any rising dust at rest-frame $4 \ \mu$m essentially rules out any significant luminosity contribution from hot ($T>500$ K) dust. Similarly, the ALMA non-detections rule out any significant contribution from cold ($T<75$ K) dust. Herschel non-detections place a ceiling of $\sim10^{12}$ $L_\odot$ in a warm component at $T\sim200$ K. Also shown is the stacked maximal LRD SED from \cite{Akins2024} (grey), scaled to the $\lambda_\mathrm{rest}=0.6 \ \mu$m luminosity of our sources, illustrating that our observations allow for a much narrower range of $L_\mathrm{IR}/L_\mathrm{optical}$ than their limits.}
    \label{fig:full_sed}
\end{figure*}

\begin{figure*}
    \centering
    \includegraphics[width=0.43\textwidth]{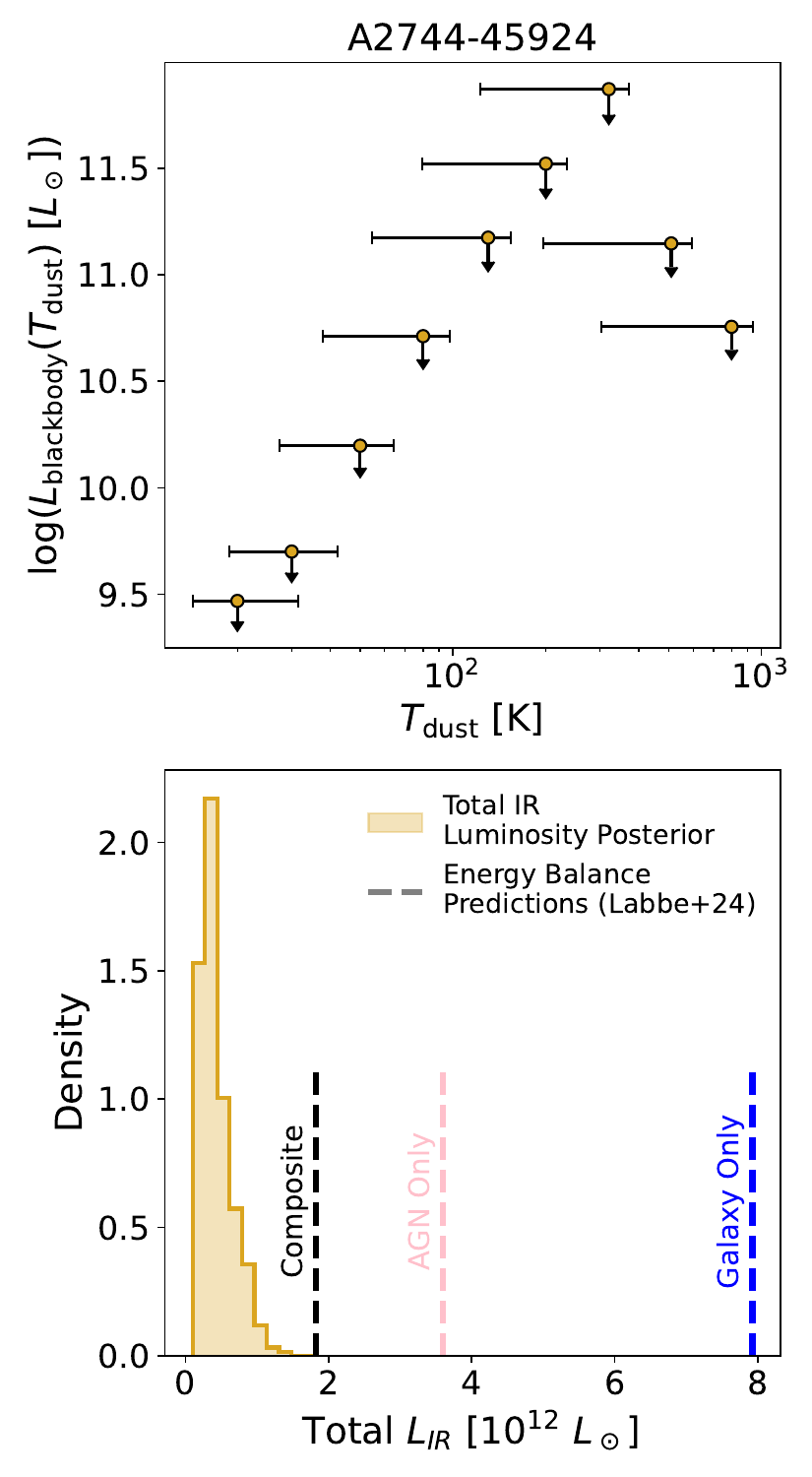}
    \includegraphics[width=0.43\textwidth]{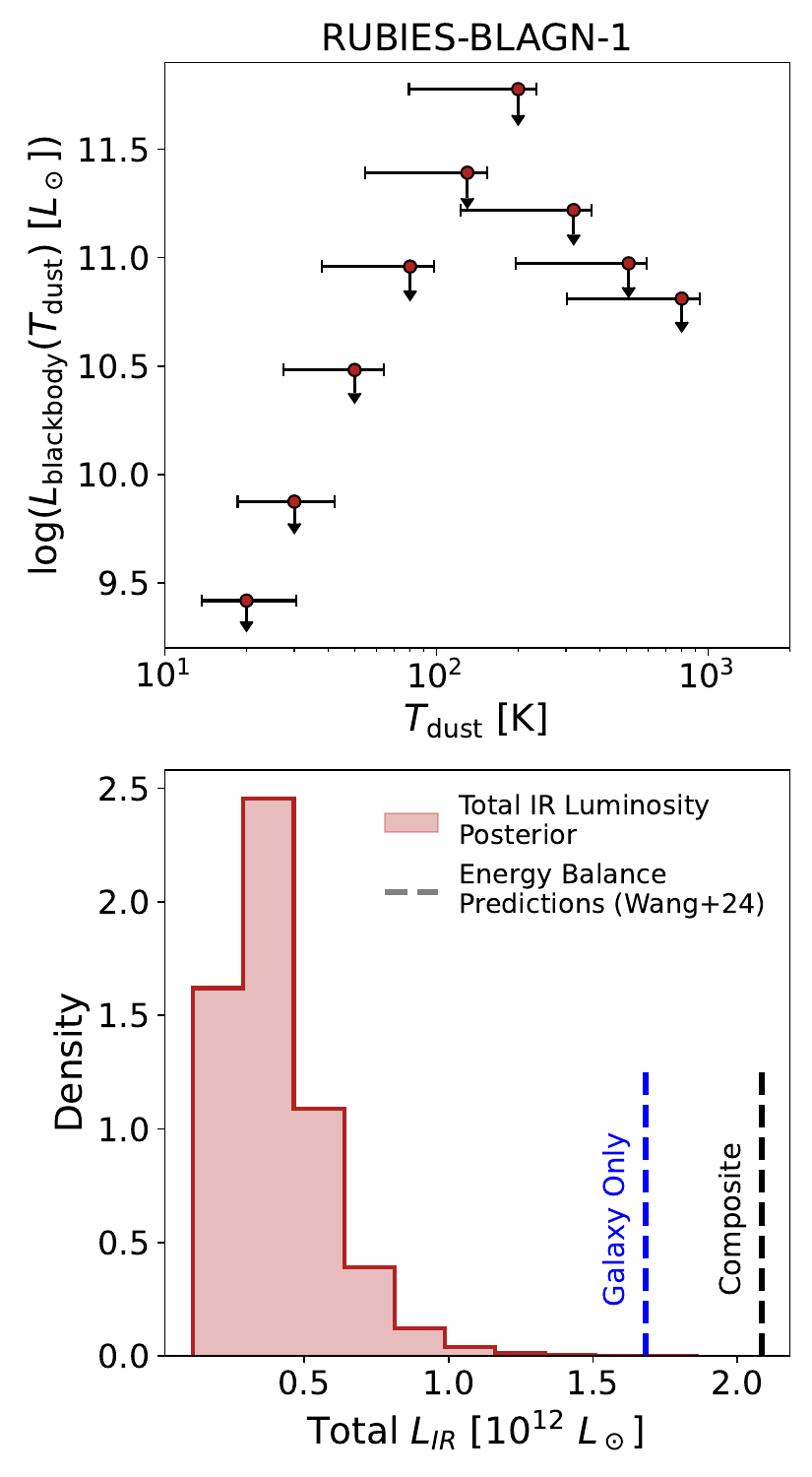}
    \caption{(Top): The $3\sigma$ upper limits of the maximum single temperature modified blackbody luminosity that can be consistent with our new IR constraints for these sources as a function of the dust temperature (see Section \ref{subsec:empirical_dust}). The width of the horizontal error bars denote the temperature range where 68\% of the power of a given blackbody is emitted. (Bottom): The posteriors for the maximum IR luminosity constrained in our fitting. We also show as dashed lines the attenuated luminosity predictions for galaxy-only (blue), and galaxy+AGN composite (black) models from \cite{Wang2024_BRD} and \cite{Labbe2024_monster}. The AGN-only energy balance prediction for A2744-45924 is also shown in pink. In both cases, the models predict IR luminosities that either exceed or are right at the edge of what is allowable by our panchromatic dust SED fitting.}
    \label{fig:LIR_limits}
\end{figure*}

\begin{deluxetable*}{ccccccccccc}[t] \label{tbl:LIR_limits}
\tablehead{
\colhead{$\log(L)$} & \colhead{20K} & \colhead{30K} & \colhead{50K} & \colhead{80K} & \colhead{130K} & \colhead{200K} & \colhead{320K} & \colhead{510K} & \colhead{800K} & \colhead{Total}
}
\tablecaption{$3\sigma$ upper limits for the FIR SED of our sources, as fit in Section \ref{subsec:empirical_dust} and illustrated in Figure \ref{fig:full_sed} and \ref{fig:LIR_limits}. For each source, we show the luminosity limit of each of the individual blackbodies included in the fitting, as well as the constraint on the total IR luminosity. Additionally, we show the same limits normalized to the continuum $\nu L_\nu$ of each of the sources at $\lambda_\mathrm{rest} = 0.6 \ \mu$m.}
\startdata
\textbf{A2744-45924} & & & & & & & & & \\
$[L/L_\odot]$ & $<9.7$ & $<9.9$ & $<10.4$ & $<10.9$ & $<11.4$ & $<11.7$ & $<12.0$ & $<11.2$ & $<10.8$ & $<12.2$ \\
$[$norm. at $0.6\mu$m$]$ & $<-1.3$ & $<-1.0$ & $<-0.5$ & $<0.0$ & $<0.5$ & $<0.8$ & $<1.1$ & $<0.3$ & $<-0.1$ & $<1.2$\\
\hline
\textbf{RUBIES-BLAGN-1} & & & & & & & & & \\
$[L/L_\odot]$ & $<9.6$ & $<10.1$ & $<10.7$ & $<11.1$ & $<11.8$ & $<12.0$ & $<11.4$ & $<11.1$ & $<10.8$ & $<12.1$ \\
$[$norm. at $0.6\mu$m$]$ & $<-1.1$ & $<-0.6$ & $<-0.0$ & $<0.4$ & $<1.0$ & $<1.3$ & $<0.7$ & $<0.3$ & $<0.1$ & $<1.4$\\
\enddata
\end{deluxetable*}

If energy balance holds and the model of an intrinsically blue source subjected to dust-reddening is correct, it is clear that the bolometric output in the FIR must emit at the specific wavelengths where our current constraints are the worst ($\lambda_\mathrm{rest}\sim 5-50 \mu$m in the case of A2744-45924, $5-150 \mu$m in the case of RUBIES-BLAGN-1). However, given the considerable systematic uncertainties and the long list of assumptions that go into both the SED fitting of LRDs and the choices of the dust SED templates, we now explore the empirical constraints on the IR luminosity, independent of any explicit predictions for the dust SED shape or normalization.

In order to do so, we construct a maximal IR SED from a series of 9 modified blackbodies (with dust emissivity $\beta=2$), logarithmically sampled between 20 K and 800 K. Using \texttt{emcee} \citep{Foreman-Mackey2013}, we fit for the luminosities of each of these blackbodies simultaneously but independently, constraining our fits with all JWST/MIRI, Herschel/PACS, and ALMA photometry at $\lambda_\mathrm{rest}>3 \ \mu$m where dust emission should dominate whatever is producing the red continuum. Non-detections are included as measurements of 0, with their uncertainties set by the $1\sigma$ flux limits. This is a conservative estimate, especially with regard to the dust continuum contribution at short wavelengths where we neglect any contribution from the LRD UV-optical continuum. In Figure \ref{fig:full_sed}, we plot the entire observed SED for our LRDs in $\nu L_\nu$, with both the observed spectrum and photometry shown. In a series of shaded curves, we show the individual contributions for each of the blackbodies for the model which produces the $3\sigma$ total IR luminosity, with the sum of these components shown as a solid line. Only warm ($T\sim200$ K) dust can produce any appreciable IR luminosity; the ALMA observations completely rule out any significant cold ($T\lesssim75$ K) dust, and the mid-IR observations similarly rule out the presence of hot ($T\gtrsim500$ K) dust. 

We additionally show the stacked maximum LRD SED from \cite{Akins2024}, rescaled to the $\lambda_\mathrm{rest}=0.6 \ mu$m luminosity of our sources, which are a factor of $\sim10$ more luminous than their typical source. We highlight that our constraints on $L_{IR}/L_{optical}$ for individual sources are significantly deeper than those in their stacks, even if the intrinsic $L_{IR}$ constraints are fairly comparable. The results of this fitting are also summarized in Table \ref{tbl:LIR_limits}, both in solar luminosity units and normalized to $\nu L_\nu$ at $\lambda_\mathrm{rest}=0.6 \ \mu$m to facilitate the use of these limits in LRDs that lack the same panchromatic SED constraints.

In the top panel of Figure \ref{fig:LIR_limits}, we collapse this maximum luminosity measurement into individual upper limits for the integrated blackbody luminosity as a function of temperature, again showing the $2\sigma$ upper limits. The x-axis error bar denotes the temperature range at which 50\% of the power in a given blackbody is emitted. In the bottom panel, we show the posterior probability distribution for the total IR luminosity that results from the combination of these curves. We additionally annotate this figure with the total $L_\mathrm{attenuated}$ values from the models discussed in Section \ref{subsec:models} and Table \ref{tbl:LIR}. All of the energy balance predictions for the emitted IR luminosity either exceed or are very near to the maximum allowed luminosity under these very conservative empirical constraints. Furthermore, the highest luminosities are only attainable if the dust essentially exclusively emits in a very narrow range of temperatures, precisely where our data is the least constraining. As such, we conclude that it is unlikely that \emph{any} of these models that rely heavily on a dust-attenuated blue component are likely to complete in their description of the physics of the LRD engine.

\subsection{\cii SFR Limits}

Independent of our energy balance analysis, the \cii non-detections in both sources can also constrain the star formation rates in these systems, under the assumption that the ISM conditions in LRDs are similar to those in typical star forming galaxies where these relations are calibrated. Under this assumption, significant star formation is also ruled out by the non-detection of \cii in both sources. Using the \cite{DeLooze2014} relation for a wide diversity of $z\sim0$ galaxies, our 3$\sigma$ SFR limits for A2744-45924 and RUBIES-BLAGN-1 are 3.4 $M_\odot\mathrm{yr^{-1}}$ and 8.1 $M_\odot\mathrm{yr^{-1}}$, respectively. If instead, we adopt the $z>6$ calibration from \cite{Schaerer2020}, the limits are 5.8 $M_\odot \mathrm{yr^{-1}}$ and 12.1 $M_\odot \, \mathrm{yr^{-1}}$. These limits are far below the star formation rates implied if one assumes the H$\alpha$ is primarily driven by star formation; even without dust correcting the H$\alpha$ luminosity, the \cite{Kennicutt1998} relation \citep[corrected to a \citealt{Chabrier2003} initial mass function following][]{Muzzin2010} yields a star formation rate of 296 $M_\odot\mathrm{yr^{-1}}$ and 156 $M_\odot\mathrm{yr^{-1}}$ for A2744-45924 and RUBIES-BLAGN-1, respectively. Star formation, whether or not it is dust obscured, must be minimal if these calibrations hold in LRDs (see also Xiao et al. submitted).

Given the significant systematic uncertainty in the masses of Little Red Dots \citep[e.g.][]{Wang2024_UB}, it is not entirely clear how to interpret these limits in the context of the evolutionary stage of the LRD host galaxy. All of the fits to A2744-45924 in \cite{Labbe2024_monster} imply masses on the order of $\log(M_\star/M_\odot)\sim11$ assuming that the red continuum is dominated by starlight, and the clustering of the system similarly implies a mass of $\log(M_\star/M_\odot)\gtrsim 10.2$. In both cases, the SFR limits measured here would imply that the host galaxies of the most luminous LRDs are decidedly below the star forming main sequence. However, in models where the AGN is responsible for the red continuum and the galaxy component is only responsible for the UV, inferred SFRs are significantly lower \citep[e.g., the composite model in][which constrains the galaxy to have $\log(M_\star/M_\odot)\sim 9.5$]{Wang2024_BRD}. The limits for RUBIES-BLAGN-1 that we report here, in conjunction with that mass, would imply a specific star formation rate of $<10^{-8.5} \mathrm{yr^{-1}}$, consistent with the fit.

\section{Discussion and Conclusions} \label{sec:discussion}

In this letter, we show that there is serious tension between the empirical IR SED limits of luminous LRDs and the bolometric luminosity that is implied by high dust attenuation in the rest-optical (under the assumption of energy balance). This tension can only barely be resolved by tuning the dust temperature distribution to emit at $T=100-300$ K, with no appreciable luminosity output in cold and hot dust reservoirs typically seen in dusty star forming galaxies and AGN, respectively. While such tightly peaked dust temperature distributions for LRDs have been proposed in \cite{Casey2024} and \cite{Li2024_LRD} \citep[and $T\sim100$ K dust has been observed in high-z dusty quasars, e.g.,][]{Fujimoto2022, FernandezAranda2025}, the fact that even such a distribution still has trouble being consistent with our data for the IR luminosities predicted by models that invoke a highly reddened blue LRD engine casts doubt on the validity of those models.

Perhaps the most natural explanation for the discrepancy between the predictions of energy balance and our FIR data is that energy balance does not hold: the attenuated luminosity inferred from the rest optical SED fitting does not equal the FIR luminosity in LRDs. However, while it is easy to imagine a scenario where energy balance \emph{under}-predicts the total IR luminosity (optically thick dust that contributes to the FIR SED while being totally invisible and unaccounted-for in the rest-optical SED fits), it is more difficult to imagine a geometric configuration where energy balance \emph{over}-predicts the IR luminosity. 

Such a geometric configuration would mean that our viewing angle is preferentially oriented such that the UV is almost entirely eclipsed, but the dust covering fraction is low enough that there is a significant range of sight lines where UV emission can escape. This would imply an analogous population of little blue dots that are comparably numerous to the LRDs studied in this work. However, LRDs are roughly a factor of $100\times$ more numerous than UV-selected quasars \citep{Greene2024}, making it difficult to imagine that LRDs are all similarly luminous quasars that are being viewed along a preferential line of sight that allows for the violation of energy balance. And if LRDs are star forming, there is no analogous UV population, as the vast-majority of UV-bright galaxies are clumpy and resolved \citep[e.g.,][]{Harikane2025}. Therefore, we proceed under the assumption that the energy balance holds.

Our tight constraints on the total IR luminosity and the dust temperature distribution cast serious doubt on the presence of a deeply dust-obscured starburst in these sources \citep[as suggested in e.g.,][]{PerezGonzalez2024_LRD, Baggen2024}. Despite their luminous H$\alpha$, they are undetected in \cii, leading to conservative limits on their SFRs of $\sim5.8$ and $12.1$ $M_\odot \mathrm{yr^{-1}}$ in A2744-45924 and RUBIES-BLAGN-1, respectively (and corroborating the growing body of evidence that LRDs are significantly \cii under-luminous relative to their Balmer emission and SFR estimates inferred from SED fitting, see Xiao et al. submitted). Furthermore, in both sources, there is essentially no room for an IR energy output that is consistent with a starburst, even if invoking highly constrained dust temperature distributions that exclusively emit at $\sim150-300$ K, where our data is the least constraining. The combination of undetected far-IR emission and non-detected \cii implies that dust-attenuated star formation cannot be contributing appreciably to the red continuum in these luminous LRDs, unless the ISM is in a configuration where the dust exclusively emits at $\sim200$ K, narrowly evading detection, and is inhospitable to \cii. 

Given that highly dust-obscured starbursts are unlikely to be consistent with the full LRD SED, it is probable that the red continuum is produced by AGN accretion disk continuum, which gives a natural explanation for the broad lines and growing evidence of variability in LRDs \citep[][]{Ji2025_LRD, Furtak2025_variability}. However, our data also disfavor any solutions where the red rest-optical continuum is the result of a dust-reddened standard AGN. The SED fitting solutions that include an AGN component also predict IR luminosities that are right at the edge of what is allowable by our limits, as attenuating a standard AGN disk to match the red continuum shape in LRDs requires a steep dust law that attenuates the entirety of the intrinsic AGN UV emission.

Ultimately, reducing the high attenuated luminosity provides the most compelling resolution to the tension between the IR predictions and the observed $L_\mathrm{IR}$ limits in LRDs. The LRD spectrum must already be redder than a standard AGN disk or young stellar population \textit{before} it sees any dust. One proposed such model is an AGN embedded in dense $T\sim10^{4}$K gas \citep{Inayoshi2024}, attenuating the UV with absorption by $n=2$ hydrogen rather than dust and naturally producing a Balmer break without invoking evolved stellar populations. However, in the implementation of such a model to fit a $z=7.3$ LRD, the suppression of the FUV by the dense gas is modest and significant attenuation ($A_V\sim2$) is still required to produce the observed red continuum shape \citep{Ji2025_LRD}. If this same approach, applied to the highly luminous LRDs in this letter, can indeed result in a significant ($\gtrsim 0.5$ dex) reduction in the attenuated luminosity, the dust SED could more comfortably fit under our IR limits without requiring quite as tight a distribution in dust temperature. It remains to be seen whether the combination of the sharp Balmer break, highly luminous broad line emission, weak UV, and lack of mid-IR and far-IR dust can be simultaneously modeled by this physical picture in these luminous sources.

There is also the possibility that the assumption of an intrinsically blue AGN or starburst that needs to be reddened by a combination of dust and gas absorption is incorrect. If, instead, the apparent spectrum of the accretion disk were significantly redder, less reddening would be needed to match the observed optical continuum. Such an SED is predicted by super-Eddington accretion,with a geometrically thick disk that emits at a lower effective temperature than a standard accretion disk \citep{Abramowicz1980, Jiang2014, Jiang2019, Madau2025}. Super-Eddington accretion models also predict weak X-ray emission \citep{Proga2005, Jiang2019}, as is seen in LRDs \citep{Lambrides2024, Yue2024, Ananna2024}. However, to date, no full SED model of super-Eddington accretion has been produced that can describe the LRD spectrum, and it remains to be seen whether an intrinsically redder AGN SED can simultaneously describe the ``V shape" and FIR limits while also accounting for the highly luminous emission lines seen in our sources.

The observations presented in this work represent the most complete SED constraints that are possible for LRDs at $z=3-5$ at present. The fact that highly luminous sources show no evidence of dust continuum emission, at any temperature, strains models that invoke a dust-attenuated blue engine to explain the rest-optical. Given that no current models appear consistent with our new constraints, these data also cast doubt on the inferred properties of LRDs, including AGN bolometric luminosities and black holes masses under the AGN interpretation or stellar masses and ages in galaxy models. Given that we do not detect LRDs in the FIR, it is unlikely that these models will yield inferred physical properties that are accurate in LRDs.

Ideally, we could make progress by drilling deeper at $\lambda_\mathrm{rest}=10-80 \ \mu$m in these systems, where our current IR constraints are weakly constrained by ancillary Herschel data. However, current telescope facilities lack the sensitivity to constrain the IR luminosity at these warm temperatures. Next-generation IR facilities like PRIMA \citep{Moullet2023} would have the required sensitivity to detect individual LRDs at these wavelengths--if they are indeed host to warm dust reservoirs. This advance would enable panchromatic SED modeling to constrain the maximum IR luminosity of LRDs beyond the upper limits that we report here. 

However, such facilities will not be online for a long time. In the mean time, lower-redshift analogues provide another path forward. 
Existing facilities like JWST/MIRI can probe hot/warm dust at $\lambda_\mathrm{rest} = 3-20 \mu$m. It is unclear whether such sources exist, and it appears that the LRD number density drops precipitously below $z\sim4$ \citep[e.g.,][]{Kocevski2024}, but it is urgent that we search large volumes given the intense level of uncertainty in the dust content of these systems.  It is clear that the current class of LRD models that assume an intrinsically blue engine (be it a starburst or a standard AGN) that is strongly attenuated by dust are at the edge of our limits for the maximum IR luminosity of LRDs. In light of this, inferred properties based on lower-redshift calibrations should be treated with skepticism until the drivers of the panchromatic SEDs of these strange luminous sources are understood.

\facilities{JWST (NIRCam, NIRSpec, MIRI), ALMA, Spitzer (MIPS), Herschel (PACs)}

\software{Astropy \citep{astropy2013, astropy2018, astropy2022}, \texttt{STScI} JWST Calibration Pipeline \citep[\url{https://jwst-pipeline.readthedocs.io/}][]{Rigby2023_jwstpipeline}, \texttt{grizli} \citep[\url{github.com/gbrammer/grizli}]{Brammer2023_grizli}, \texttt{msaexp} \citep[\url{https://github.com/gbrammer/msaexp}][]{Brammer2023_msaexp},
Matplotlib \citep{Hunter:2007}, Flexible Stellar Population Synthesis \citep{Conroy2009, Conroy2010}, SEDPy \citep{sedpy2019}}

\begin{acknowledgements}

Support for this work was provided by The Brinson Foundation through a Brinson Prize Fellowship grant. DS acknowledges Zhengrong Li for kindly sharing model dust SEDs, Tim Rawle for helping with accessing archival Herschel Lensing Survey data, and Xiaohui Fan for helpful conversations that steered the direction of this work. Support for this work was provided by NSF/AAG \#2306950. Support for this work for RPN was provided by NASA through the NASA Hubble Fellowship grant HST-HF2-51515.001-A awarded by the Space Telescope Science Institute, which is operated by the Association of Universities for Research in Astronomy, Incorporated, under NASA contract NAS5-26555. This work has received funding from the Swiss State Secretariat for Education, Research and Innovation (SERI) under contract number MB22.00072, as well as from the Swiss National Science Foundation (SNSF) through project grant 200020\_207349. The Cosmic Dawn Center is funded by the Danish National Research Foundation under grant DNRF140. AZ acknowledges support by Grant No. 2020750 from the United States-Israel Binational Science Foundation (BSF) and Grant No. 2109066 from the United States National Science Foundation (NSF); and by the Israel Science Foundation Grant No. 864/23. The work of CCW is supported by NOIRLab, which is managed by the Association of Universities for Research in Astronomy (AURA) under a cooperative agreement with the National Science Foundation.  SA acknowledges sup-
port from the JWST Mid- Infrared Instrument (MIRI) Science Team Lead, grant 80NSSC18K0555, from NASA Goddard Space Flight Center to the University of Arizona.

\end{acknowledgements}

\bibliography{Bright_LRD_FIR}

\end{document}